\documentclass[twocolumn,prb,showpacs,floatfix]{revtex4}
\usepackage{graphicx}
\usepackage{dcolumn}
\usepackage{bm}
\usepackage{amssymb}
\usepackage{amsmath}
\usepackage{subfigure}
\usepackage{amsfonts}
\usepackage{xcolor}
\usepackage{psfrag}
\usepackage{float}
\usepackage{multirow}
\providecommand{\U}[1]{\protect\rule{.1in}{.1in}}

\newcommand{\comp}{(BEDT-TTF)$_2$Ag(CF$_3$)$_4$(TCE)}

\newcommand{\dual}{$\kappa$-$\alpha_1'$-(BEDT-TTF)$_2$Ag(CF$_3$)$_4$(TCE)}

\newcommand{\ag}{[Ag(CF$_3$)$_4$]$^{-}$}
\newcommand{\one}{$\kappa$-$\alpha_1'$}
\newcommand{\two}{$\kappa$-$\alpha_2'$}

\begin{document}
\title{Role of layer packing for the electronic properties
of  the organic superconductor (BEDT-TTF)$_2$Ag(CF$_3$)$_4$(TCE)} 
\author{Michaela Altmeyer}
\email{altmeyer@itp.uni-frankfurt.de}
\affiliation{Institut f\"ur Theoretische Physik, Goethe-Universit\"at Frankfurt, 60438
Frankfurt am Main, Germany}
\author{Roser Valent\'i}
\affiliation{Institut f\"ur Theoretische Physik, Goethe-Universit\"at Frankfurt, 60438
Frankfurt am Main, Germany}
\author{Harald O. Jeschke}
\affiliation{Institut f\"ur Theoretische Physik, Goethe-Universit\"at Frankfurt, 60438
Frankfurt am Main, Germany}
\date{\today}
\pacs{74.70.Kn, 71.20.Rv, 71.10.Fd, 71.15.Mb}

\begin{abstract}
  The charge-transfer compound (BEDT-TTF)$_2$Ag(CF$_3$)$_4$(TCE)
  crystallizes in three polymorphs with different alternating layers:
  While a phase with a $\kappa$ packing motif has a low
  superconducting transition temperature of $T_c=2.6$~K, two phases
  with higher $T_c$ of $9.5$ and $11$~K are multi-layered structures
  consisting of $\alpha'$ and $\kappa$ layers. We investigate these
  three systems within density functional theory and find that the
  $\alpha'$ layer shows different degrees of charge order
  for the two $\kappa$-$\alpha'$ systems and directly influences the
  electronic behavior of the conducting $\kappa$ layer. We discuss the
  origin of the distinct behavior of the three polymorphs and propose
  a minimal tight-binding Hamiltonian for the description of these
  systems based on projective molecular Wannier functions.
\end{abstract}

\maketitle

\hspace{5.2in}

\section{Introduction}
For a few decades organic charge-transfer (CT) salts built of donor
and acceptor molecular complexes have attracted a lot of attention due
to the variety of ground states in their phase
diagrams~\cite{Jerome1991,Mori2006,Manna2010,Kanoda2011,Jacko2013}.
Application of external or chemical pressure can lead to
antiferromagnetic insulating, charge ordered, spin-density wave, spin
liquid, or unconventional superconducting ground states.  Tendencies in
the dimensionality of the electronic transport are often determined by
the choice of conducting molecules: Compounds containing TMTTF
(tetramethyltetrathiafulvalene) molecules, for example, are typically
one-dimensional~\cite{Jerome1991,Jacko2013}, whereas several phases of
BEDT-TTF (bisethylenedithio-tetrathiafulvalene) based salts show
two-dimensional behavior.  However, the arrangement of the (donor)
molecules in these complexes is decisive.  Among the BEDT-TTF family
of CT salts, many different packings classified as $\alpha$,
$\alpha'$, $\beta$, $\beta'$, $\beta''$, $\delta$, $\kappa$, and
$\theta$ have been experimentally realized and a wide range of
different physical properties was
found~\cite{Mori1998,Mori1998b,Mori1999,Powell2006}.  Depending on the
preparation conditions, different polymorphs of one structure can be
synthesized; for example, (BEDT-TTF)$_2$I$_3$ crystallizes in $\alpha$,
$\beta$, $\theta$, and $\kappa$ forms~\cite{Yoshimoto1999}.
Polymorphs provide an opportunity to explore the influence of the
packing motif on the electronic properties. Effects originating from
differences in the anion layer composition can be excluded in this
case since this
layer remains unaltered in the polymorph family.

Here we consider the polymorph charge-transfer salt family {\comp}
(see Fig.~\ref{fig:structure}) first synthesized by Schlueter and
collaborators~\cite{Schlueter1994}.
TCE stands for 1,1,2-trichloroethane and in the following we will make
use of the common abbreviation ET for BEDT-TTF. These systems show a
metallic behavior at low temperatures and a $T_c$ of $2.6$~K to a
superconducting state was measured for the single-layered compound
(Fig.~\ref{fig:structure} (a)) where the ET molecules form dimers
arranged in a so-called $\kappa$ pattern.  The term $\kappa_L$
phase was coined for this structure, with the index $L$ referring to
the low $T_c$.
Structural refinement of the other two multiphase
polymorphs~\cite{Schlueter2010,Kawamoto2012}
(Figs.~\ref{fig:structure} (b), (c) and
Fig.~\ref{fig:structure_overview}) showed the presence of
charge-ordered layers in $\alpha'$ packing between the $\kappa$-type
layers; $\alpha'$ phases have also been characterized as Mott-Hubbard
insulators~\cite{Obertelli1989,Parker1989,Kubo2012}. Even though as insulating
layers they do not contribute directly to superconductivity, their
existence seems to enhance the superconducting transition temperature
in {\comp}.  While the dual-layered {\one} compound exhibits
superconductivity at a critical temperature of $9.5$~K, which is
approximately 3.5 times higher than the $T_c$ of the $\kappa_L$ phase,
the four-layered {\two} phase shows superconductivity at $11$~K and
therefore belongs to the organic superconductors with the highest
measured critical temperatures.

In this work we perform density functional theory (DFT) calculations for
 the three polymorphs. We especially focus on the effects
of the $\alpha'$ layers on the electronic properties of the $\kappa$
layers in the dual and four-layered systems and perform a comparative analysis
of the three systems in terms of {\it ab initio} derived tight-binding Hamiltonians
using the projective Wannier method. 
While all three systems show apparently similar $\kappa$ bands, the charge
ordering in the $\alpha'$ layer in {\one} and {\two} 
influences significantly the magnitude of the 
hoppings in the conducting $\kappa$ layer. Analysis of the degree
of frustration within a minimal triangular lattice model hints to
the different superconducting $T_c$ in these systems.

\section{Crystal structure}

\begin{figure}
\centering
\includegraphics[width=\columnwidth]{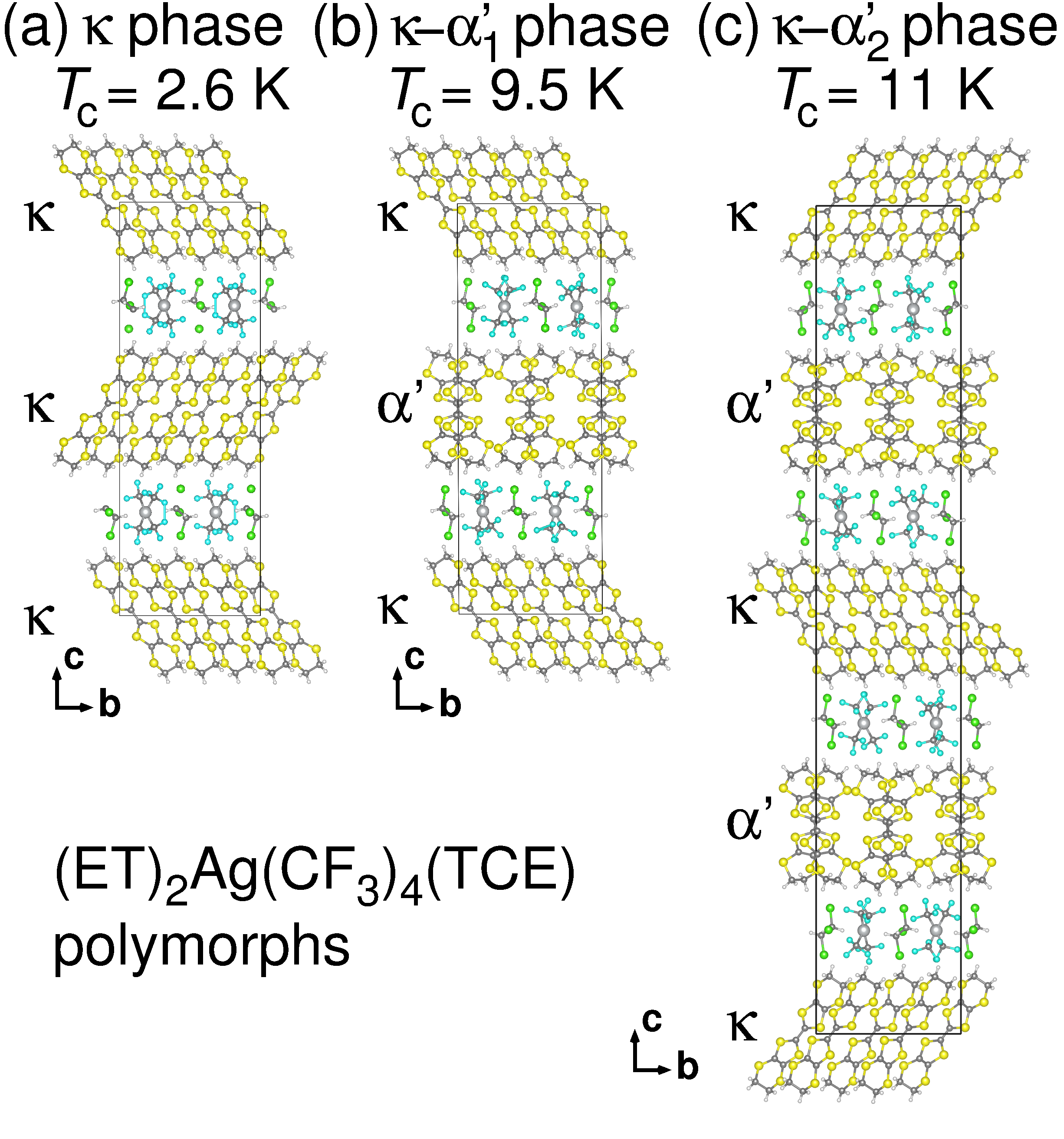}
\caption{Crystal structures of the three polymorphs of {\comp}.
}
\label{fig:structure}
\end{figure}

\begin{figure}
\centering
\includegraphics[width=\columnwidth]{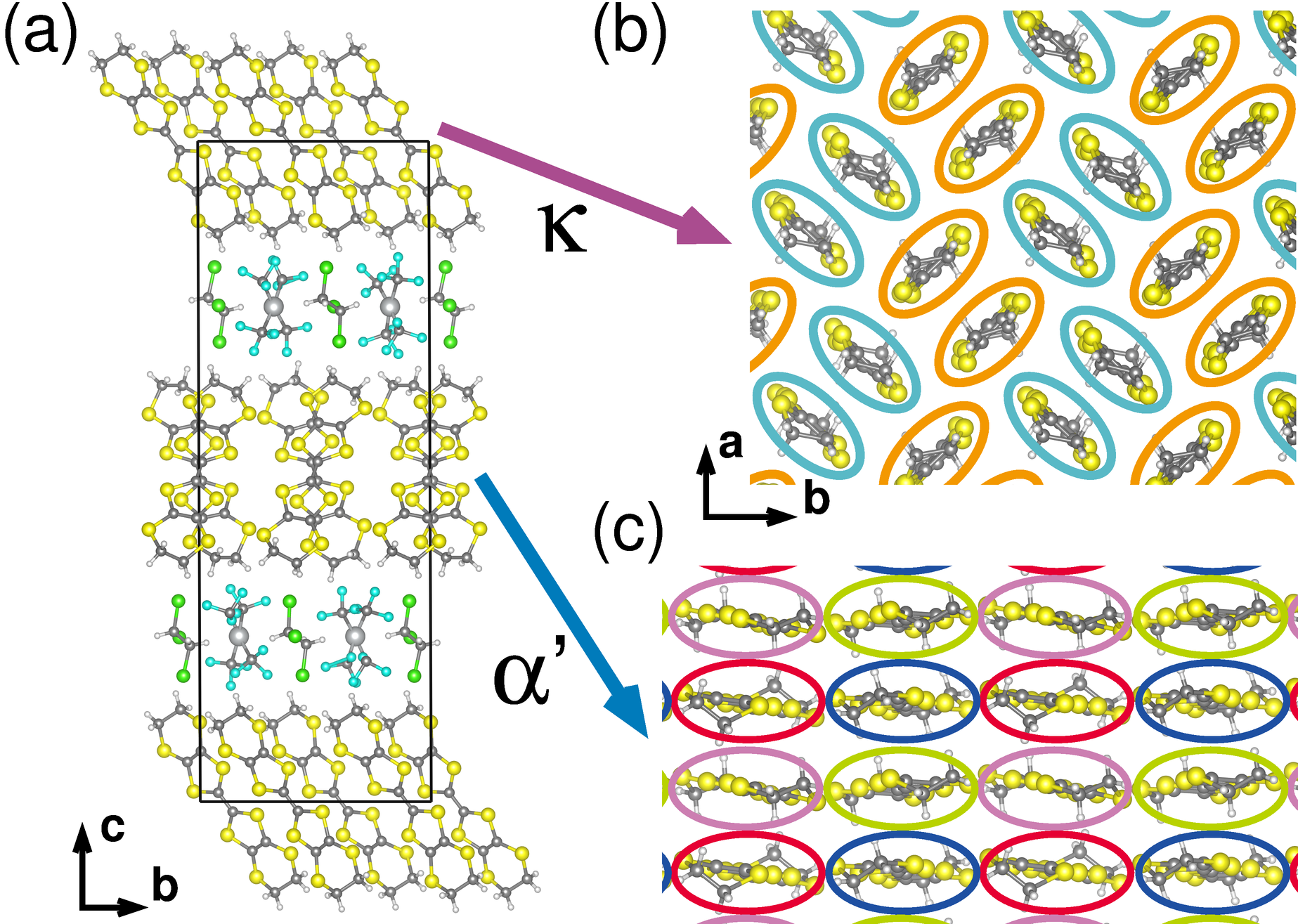}
\caption{Crystal structure of {\dual} together with cuts illustrating
  the two alternating ET patterns. In (b) and (c), colored outlines
  mark symmetry inequivalent ET molecules. }
\label{fig:structure_overview}
\end{figure}

\begin{figure}
\centering
\includegraphics[width=\columnwidth]{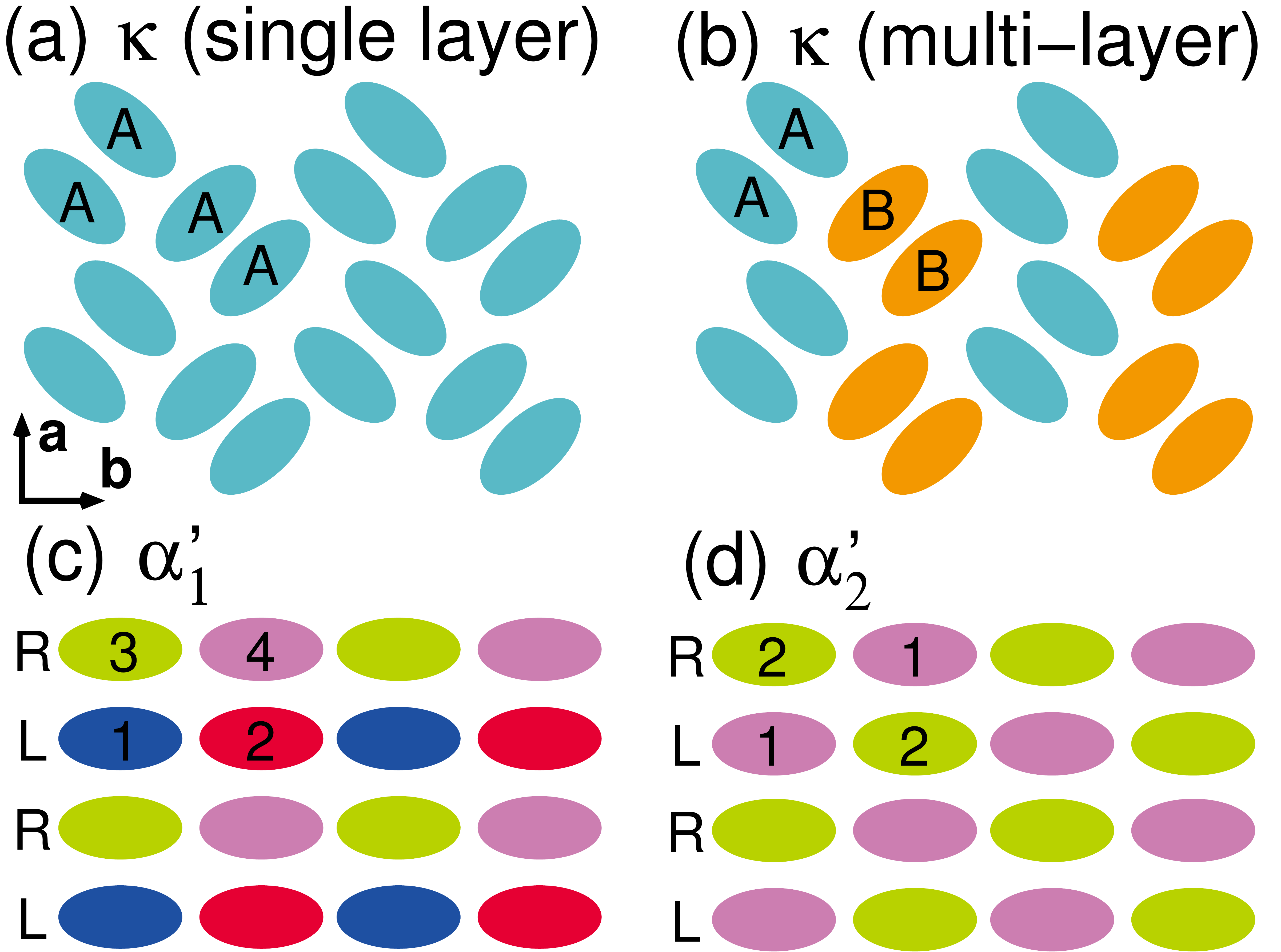}
\caption{Schematic of BEDT-TTF molecule arrangement in the considered
  materials. (a) $\kappa$-type packing in the low $T_c$ {\comp}
  compound. (b) $\kappa$-type packing in the two high $T_c$
  compounds. $\alpha_1'$-type packing in the two-layer compound. (c)
  $\alpha_2'$-type packing in the four-layer compound. Symmetry
  inequivalent molecules are marked by numbers and colored
  differently. R (L) indicate tilting of the molecules to the right
  (left) in the $\alpha'$ layers.}
\label{fig:layerscheme}
\end{figure}

{\bf $\kappa$ phase.-} The single-layered phase crystallizes in the
orthorhombic $P\,nma$ space group and its 
 unit cell  contains two donor
layers which are separated by an insulating layer consisting of the
anion {\ag} and the solvent TCE~\cite{Geiser1995}. However, the anion
layer is disordered. In order to simplify the density functional
theory calculations we choose one of the two symmetry-allowed orientations
of the anion
and lower the symmetry to $P\,2_1/c$;
the corresponding simplified structure is shown in
Fig.~\ref{fig:structure}~(a). Note that for better comparability with
the {\one} and {\two} phases we denote with $a$ the short in-plane
axis, with $b$ the long in-plane axis and with $c$ the stacking
direction (even though the original $P\,nma$ structure has a $b$
stacking direction).  In the two donor layers, the ET molecules are in
the $\kappa$ packing motif (see  Fig.~\ref{fig:layerscheme}~(a)) but
with alternating tilt of the molecules with respect to the acceptor
layer.

{\bf {\one} phase.-} The triclinic crystal structure $P\,\bar{1}$ of
the dual-layered compound is shown in Fig.~\ref{fig:structure}~(b).
Compared to the single-layered phase, the layer in the center of the
unit cell is here replaced by an $\alpha'$ packed layer.  This packing
motif is characterized by molecules that are lined up on a rectangular
lattice (see Fig.~\ref{fig:structure_overview} and
Fig.~\ref{fig:layerscheme}~(c)). The rows of molecules along $b$
alternate between a right and a left tilt with respect to the $a$
axis. This leads to the characteristic cross pattern when viewed along
$a$ (Fig.~\ref{fig:structure_overview} and
Fig.~\ref{fig:layerscheme}~(c)).  In contrast to other $\alpha'$
packed structures~\cite{Mori1999}, no dimerization and therefore also
no shift along the long unit cell axis has been observed.  The low
symmetry of the space group means that in the $\kappa$ layer, the two
ET molecules in a dimer are still related by inversion symmetry but there
are two symmetry inequivalent dimers, and in the $\alpha'$ layer there
are even four inequivalent ET molecules (see
Fig.~\ref{fig:structure_overview} and Figs.~\ref{fig:layerscheme}~(b)
and (c)).

{\bf {\two} phase.-} The polymorph with the highest superconducting
critical
temperature has the largest unit
cell~\cite{Kawamoto2012,Kawamoto2013,alpha2primestructure}, where two $\kappa$
layers alternate with two $\alpha'$ layers
(Fig.~\ref{fig:structure}~(c)). Due to the monoclinic $P\,2_1/n$
symmetry (space group No. 14), every second $\kappa$($\alpha'$) layer
is shifted by half the lattice vector $a$. Note that DFT codes usually
allow only one setting of space group No. 14; we therefore perform the
calculations in a $P\,2_1/c$ setting.  As in the case of {\one}, the
monoclinic space group leads to two symmetry inequivalent (ET)$_2$
dimers in the $\kappa$ layer; the two ET molecules in a dimer are
related by inversion symmetry (see Fig. \ref{fig:layerscheme} (b)).
The symmetry also leads to a checkerboard pattern of symmetry related
ET molecules in the $\alpha'$ layer as displayed in
Fig. \ref{fig:layerscheme} (d).

\section{Electronic structure}

\begin{figure}
\centering
\includegraphics[width=\columnwidth]{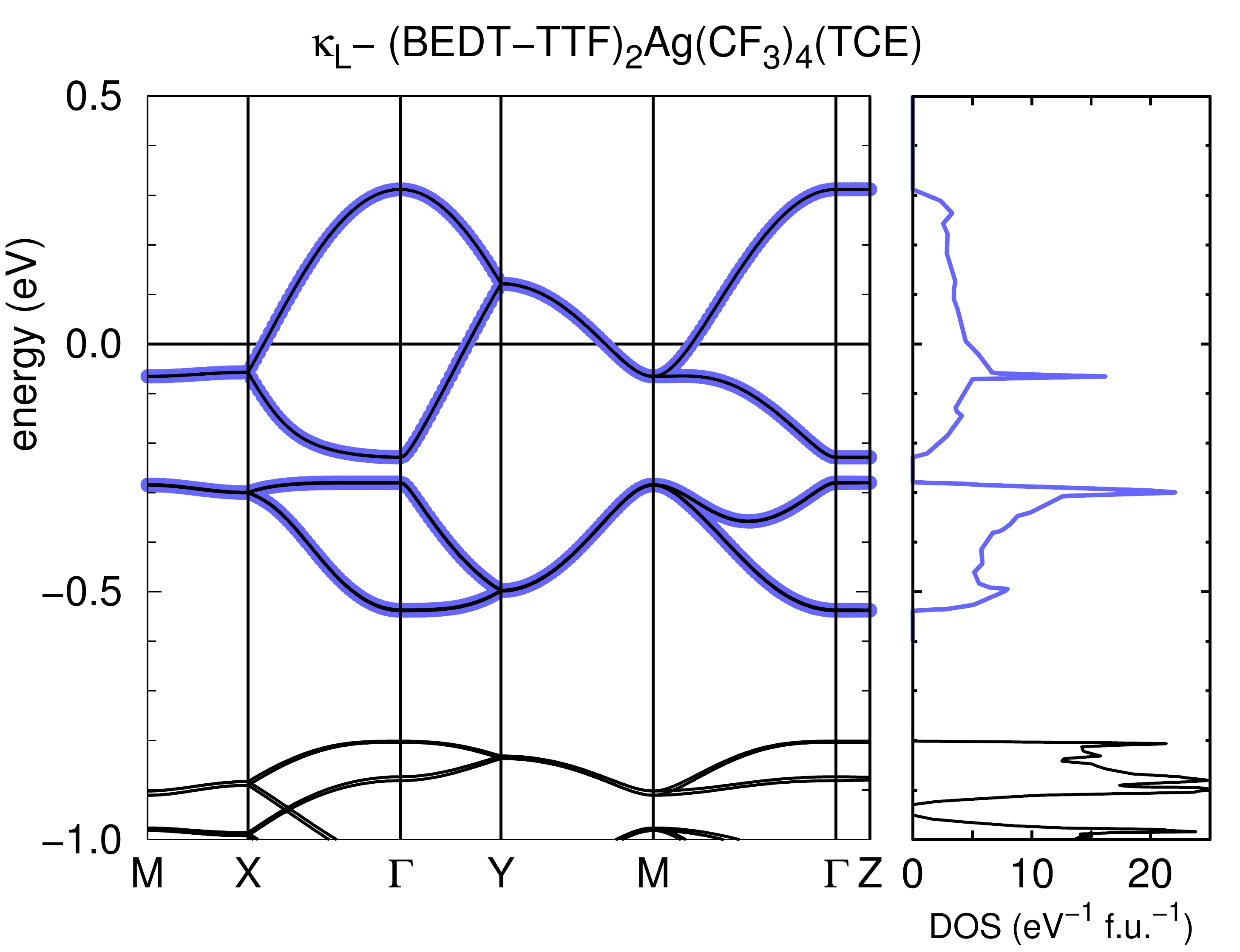}
\caption{DFT band structure and density of states of the $\kappa$
  phase. The blue bands are determined by the projective Wannier function method and originate from the ET HOMO orbitals. }
 \label{fig:kappaband}
\end{figure}

\begin{figure}[t]
\centering
\includegraphics[width=\columnwidth]{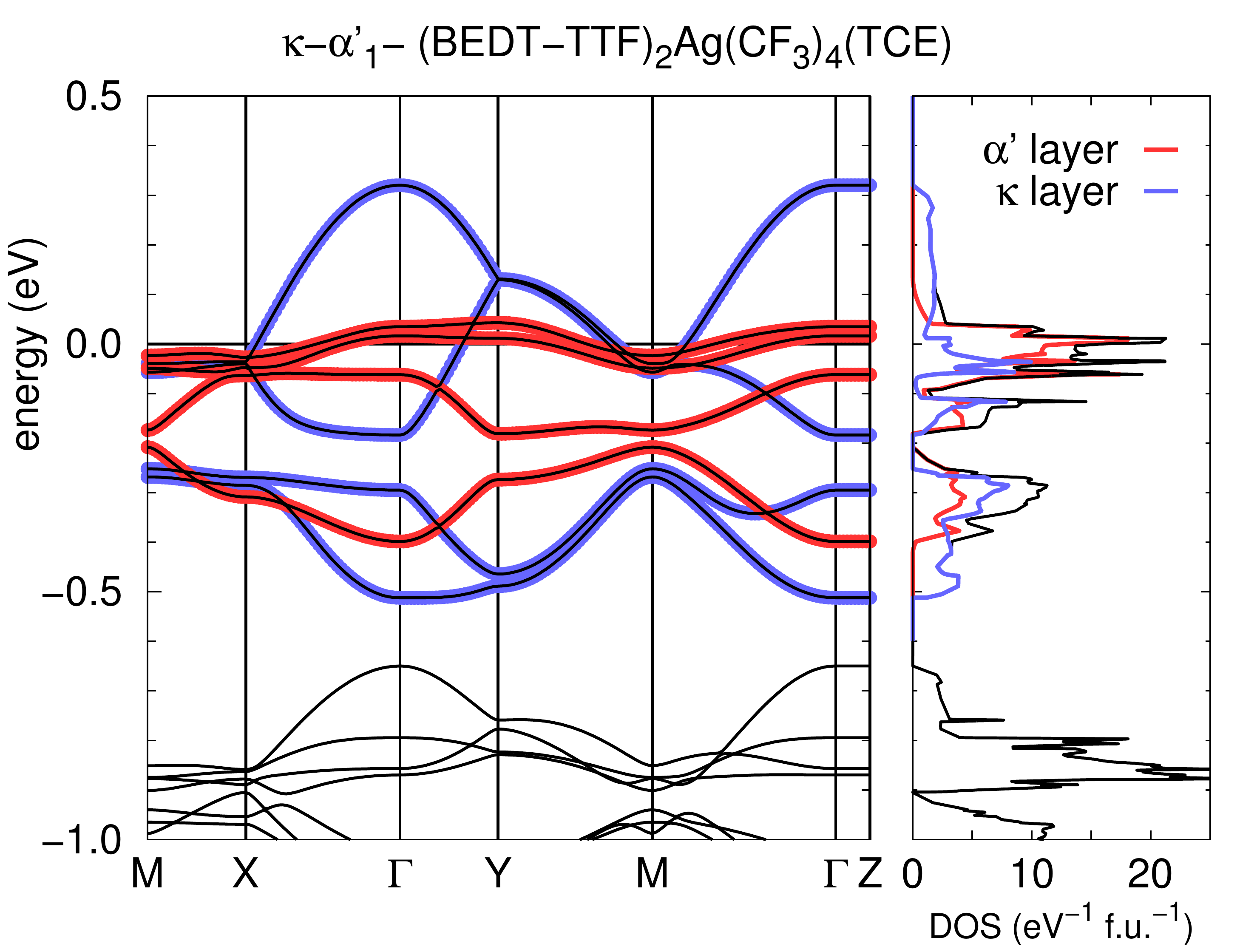}
\caption{ 
Band structure and density of states (DOS) of the {\one} phase. The Wannier bands (blue symbols for $\kappa$ orbitals and red symbols for $\alpha'$ orbitals) are in excellent agreement with the DFT bands.}
\label{fig:alpha1band}
\end{figure}

\begin{figure}[t]
\centering
\includegraphics[width=0.9\columnwidth]{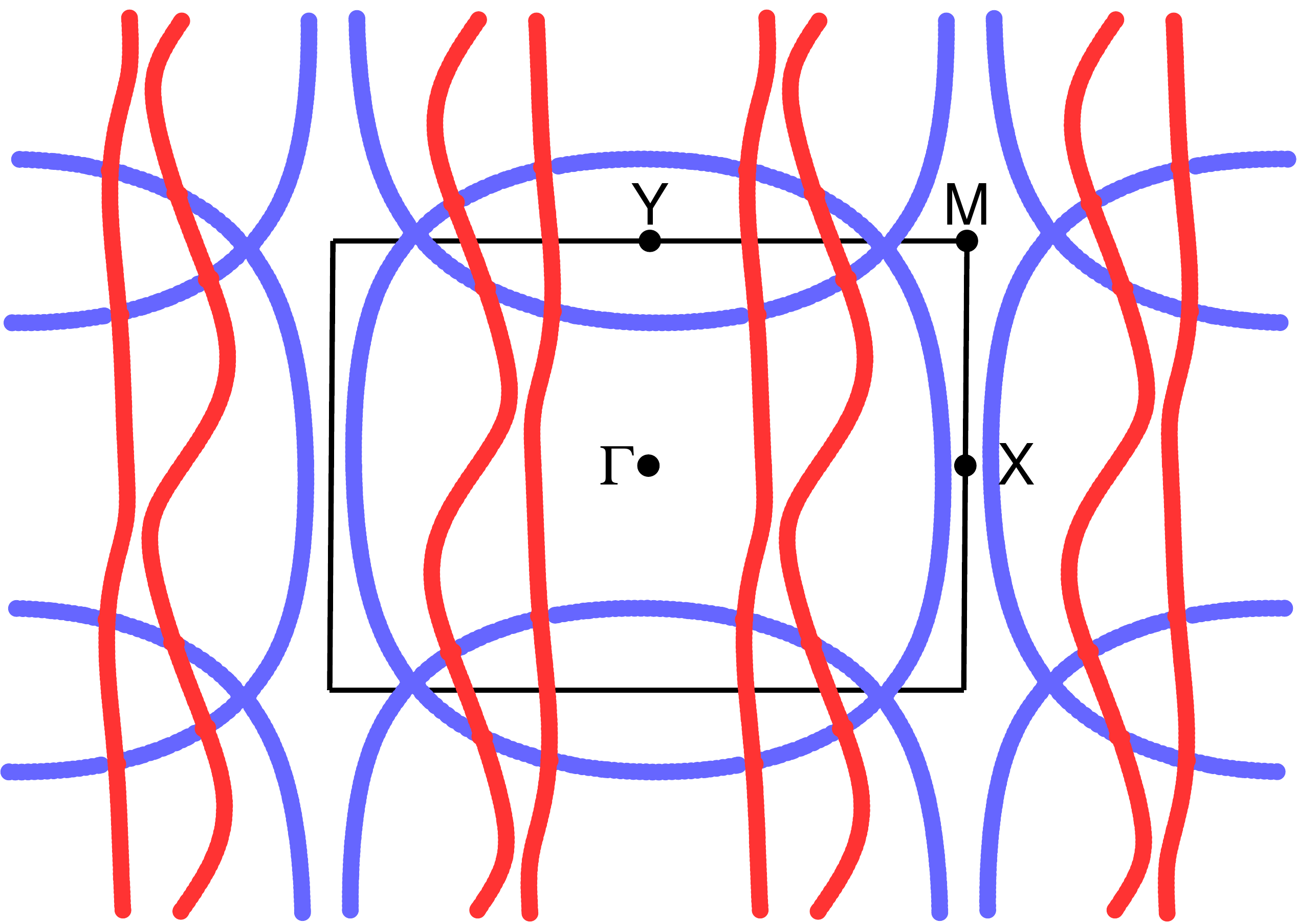}   
\caption{Calculated Fermi surface of the dual layered compound: The
  elliptical shapes in blue are due to the bands arising from the ET
  molecules of the $\kappa$-layer, while the wiggly lines in red are the
  almost one-dimensional features resulting from the $\alpha'$-layer.}
\label{fig:alpha1FS}
\end{figure}

\begin{figure}
\centering
\includegraphics[width=\columnwidth]{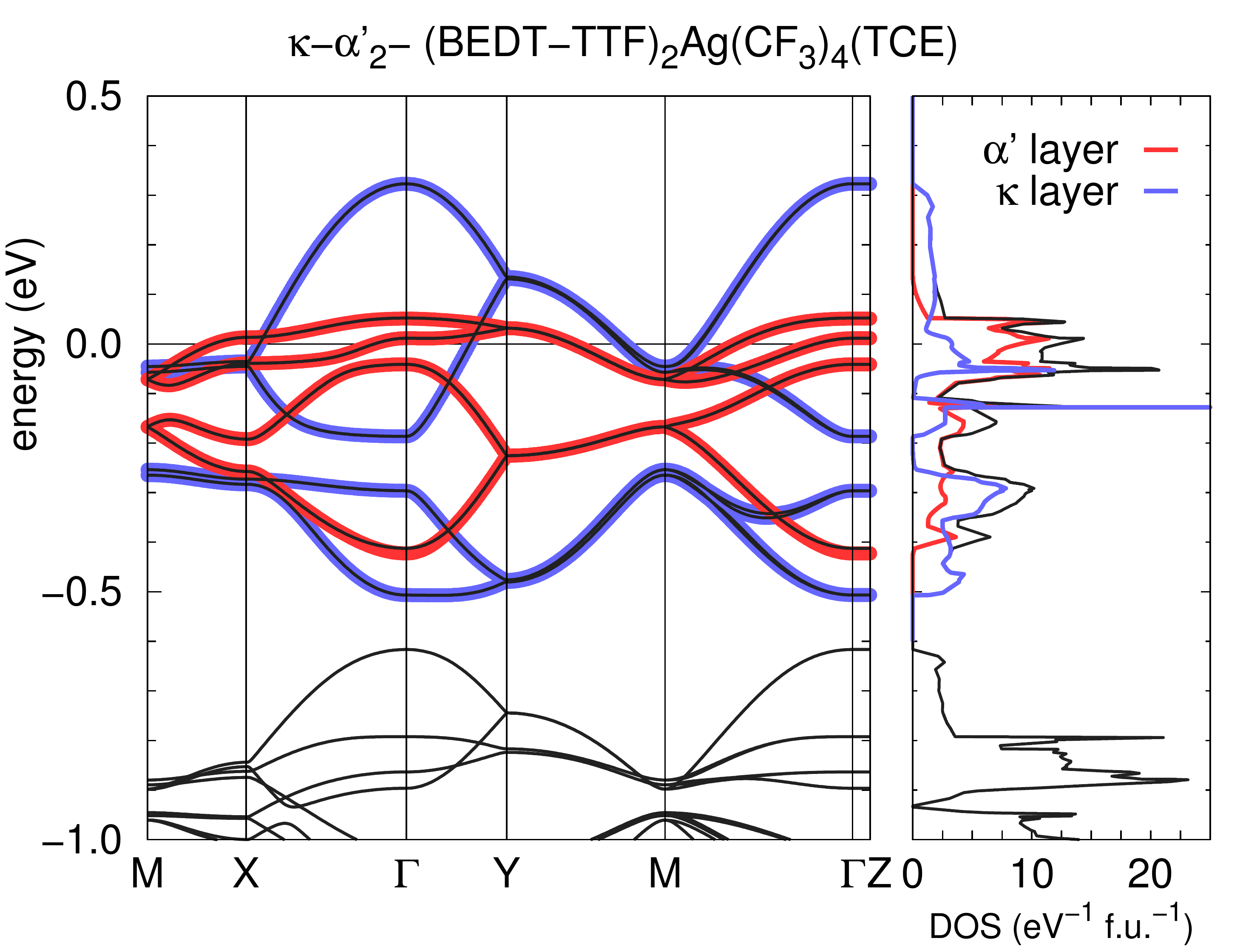}
\caption{Band structure and density of states of the {\two}
  phase. The DFT band structure is shown with black lines, while the Wannier bands are displayed with blue ($\kappa$) and red ($\alpha'$) symbols.}
\label{fig:alpha2band}
\end{figure}

We determine the electronic structure of {\comp} using the all
electron full potential local orbital basis as implemented in the FPLO
code~\cite{Koepernik1999} and the generalized gradient approximation
functional~\cite{Perdew1996}. A $6 \times 6 \times 6$ $k$ mesh was
employed to converge the energy and charge density.

{\bf $\kappa$ phase.-} 
The band structure and density of states (DOS) of the $\kappa$ phase
are presented in Fig.~\ref{fig:kappaband}. There are eight bands in the
energy window $[-0.6 \,{\rm eV}, 0.4 \,{\rm eV}]$ deriving exclusively
from the highest occupied molecular orbitals (HOMOs) of the eight ET
molecules in the unit cell. As there is only a very small dispersion
in the stacking direction ($\Gamma$-$Z$ path) the bands originating
from the two layers in the unit cell
are almost degenerate. The dimerization of the ET
molecule pairs is reflected in the bonding ($[-0.6 \,{\rm eV},-0.3
\,{\rm eV}]$) and antibonding ($[-0.3 \,{\rm eV},0.4 \,{\rm eV}]$)
character of the bands.
The bandstructure of $\kappa$-{\comp} is very similar to $\kappa$-(BEDT-TTF)$_2X$
compounds with other anions $X$~\cite{Kandpal2009,Nakamura2009}. 
Integrating the partial densities of states for symmetry inequivalent
molecules in the range $[-0.6,0]$~eV results in a charge transfer
of one hole per (ET)$_2$ dimer to the anion layer.

{\bf {\one} phase.-} In Fig.~\ref{fig:alpha1band} we show the band
structure and layer-resolved density of states of the dual-layered
compound. The anion layer has no weight at the Fermi level and
therefore does not contribute to the electronic transport.  
The bands originating from the $\kappa$ layer are similar in shape to
the bands of the $\kappa$ phase (Fig.~\ref{fig:kappaband}) while the
$\alpha'$ bands are narrower and less dispersive than the $\kappa$
bands.  The corresponding Fermi surface of the {\one} phase is shown
in Fig.~\ref{fig:alpha1FS}. Here, the elliptical shape corresponds to
the $\kappa$ layer and is typical for this packing
motif~\cite{Caulfield1994,Kandpal2009}.  In contrast, the $\alpha'$
Fermi surface (wiggly lines in Fig.~\ref{fig:alpha1FS}) is
quasi-one-dimensional and has no dispersion in the $k_z$ direction.

Recent de Haas-van Alphen experiments for the dual-layered compound
observed only elliptic orbitals originating from the $\kappa$
bands~\cite{alpha2primestructure} and, as expected, no $\alpha'$ bands were
detected.  However, one should mention that on the one hand the
$\alpha'$ packed systems have been reported in the past as
Mott-Hubbard insulators~\cite{Obertelli1989,Parker1989, Kubo2012} and,
on the other hand, DFT underestimates correlation effects and cannot
reproduce the insulating behavior of a Mott system.  A better
treatment of correlations beyond DFT in organic
materials~\cite{Ferber2014} (presently beyond the scope of this study)
could lead to an opening of a gap at the Fermi level in the $\alpha'$
bands and to a Fermi surface with only $\kappa$ bands. Note, however,
that direct hybridization between $\alpha'$ and $\kappa$ layers is
almost negligible (maximum $\alpha'$ to $\kappa$ hopping parameters in
{\one} and {\two} are 0.2-0.3~meV) so that it is justified
to focus our investigation on the properties of the $\kappa$ layers
even in {\one} and {\two} as these layers will be responsible for the
observed superconductivity.

\begin{figure}
\centering
\includegraphics[width=0.75\columnwidth]{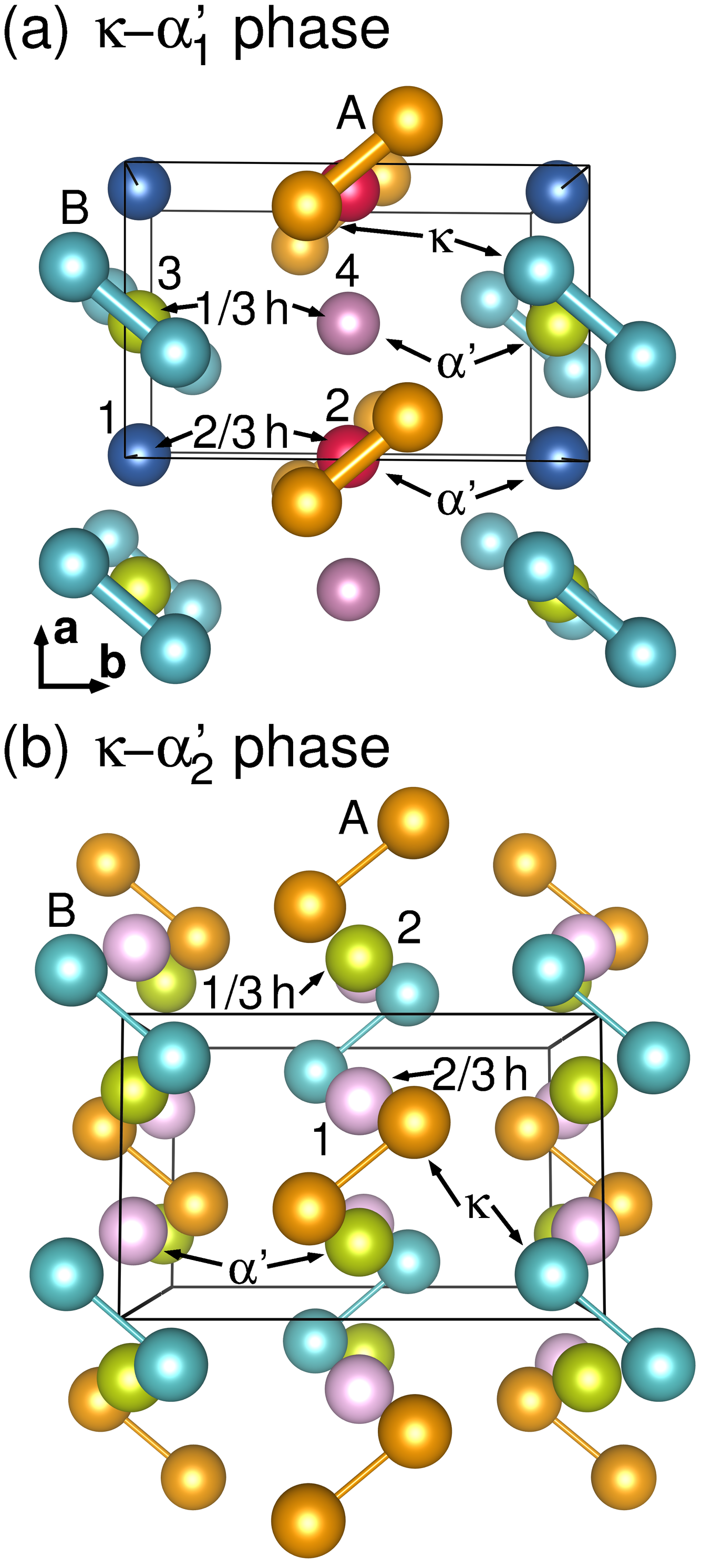}
\caption{Illustration of the two different $\kappa$ dimers in the
  charge ordered environment as created by the $\alpha'$ layer in the
  {\one} and {\two} phase. ET molecules are shown as spheres; the two 
  symmetry inequivalent {$\kappa$} layer ET molecules are labeled A and B.
  (a) The four inequivalent {$\alpha'$} layer ET molecules in the {\one} 
  phase carry numbers 1 through 4. (b) There are only two inequivalent
  {$\alpha'$} layer molecules in the {\two} phase that are labeled 1 and 2.}
\label{fig:scheme}
\end{figure}

The charge transfer in this system has more features than in the pure
$\kappa$ phase compound due to the low symmetry and the presence of
the $\alpha'$ layer.  We find that the right tilted ET molecules in
the $\alpha'$ plane contribute a charge of the order of 1/3 electron
(0.343 for ET molecule 3 as denoted in Fig.~\ref{fig:layerscheme}~(c)
and 0.316 for molecule 4). The other two left tilted molecules show a
larger charge transfer of the order of 2/3 of an electron (0.658 for
molecule 1 and 0.705 for molecule 2).  For clarity, we depict in
Fig.~\ref{fig:scheme}~(a) this charge ordering.  A similar charge
disproportionation has already been observed in other $\alpha'$ charge
transfer salts (e.g. in $\alpha'$-ET$_2$Ag(CN)$_2$~\cite{Guionneau1995}), 
where the homogeneous charge transfer of the
ET molecules at high temperature is redistributed to a 1/3 versus 2/3
order upon cooling.

In the $\kappa$ layer, the symmetry of the crystal leads to two
symmetry inequivalent (ET)$_2$ dimers with distances between the ET
molecules in a dimer of $d_{A}=3.74~\textnormal{\AA}$ and
$d_{B}= 3.77~\textnormal{\AA}$.  However, the charge transfer from
these dimers - 0.493 electrons for ET molecule A (see
Fig.~\ref{fig:layerscheme}~(b)) and 0.494 for molecule B - to the anion 
layer is the same within the computational accuracy.

{\bf {\two} phase.-} Fig. \ref{fig:alpha2band} shows the band
structure and density of states of the four-layered compound. As the
unit cell consists of four donor layers, there are 16 bands
corresponding to the HOMOs of the ET molecules. However, as the
dispersion along the stacking direction is extremely small, the bands
originating from the two identically packed layers in the unit cell
are nearly pairwise degenerate and the band structure is very similar
to that of the {\one} compound (see Appendix B).
 Nevertheless, the subtle quantitative
differences between the electronic structure of the two systems, in
particular in the $\alpha'$ bands, have important consequences on the
behavior of the materials as it is reflected, for instance, in the
charge transfer.  The $P\,2_1/c$ symmetry in {\two} poses stronger
restrictions on the charge transfer in the $\alpha'$ layer compared to
the $P\,\bar{1}$ {\one} system. The symmetry equivalent ET molecules
form a checkerboard pattern on the rectangular lattice as shown in
Fig.~\ref{fig:layerscheme}~(d). The ET molecules on one of the
$\alpha'$ sublattices donate 0.367 electrons, while the other
molecules transfer approximately 0.611 electrons. Thus, the average
charge transfer from the $\alpha'$ layers to the anion layer is
slightly less than half an electron per ET molecule.

As in the {\one} system, there are two symmetry inequivalent (ET)$_2$ dimers
in the $\kappa$ layer. The distances of the molecules within the
dimers differ slightly, at $d_A=3.73$~{\AA} and $d_B= 3.76$~{\AA}.
There is a small charge disproportionation between the dimers, with ET
molecules of dimer A transferring 0.520 electrons to the anion, while
those of dimer B transfer 0.500 electrons (compare
Fig.~\ref{fig:layerscheme}~(b)). Thus, the $\kappa$ layer compensates
for the slightly too low charge transfer of the $\alpha'$ layer.

An important difference between the {\one} and {\two} phases is that
the constraint imposed by the higher symmetry in {\two} translates
into a more symmetric charge order in the $\alpha'$ layer in {\two}
than in {\one} and the corresponding $\alpha'$-$\kappa$ stacking in
{\two} shows the center of the (ET)$_2$ dimers of the $\kappa$ layers 
always aligned between the two differently charged ET molecules in the 
$\alpha'$ layer (Fig.~\ref{fig:scheme}~(b)), contrary to what happens in {\one} . 
Note, that the further inclusion of correlations may change the degree of 
charge order while the symmetry constraints will keep the pattern the same.
This charge arrangement has important consequences on the Hamiltonian 
description of these systems as we show below.

\section{Tight-binding Hamiltonian}\label{TB Ham}

We use the projective Wannier function method as implemented in
FPLO~\cite{Eschrig2009} in order to obtain the tight-binding
parameters for the conducting $\kappa$ layers in {\comp} from Wannier
function overlaps.  Considering the central point of the inner C-C bond
of the ET molecules as sites, the tight-binding (TB) Hamiltonian can be written as:
\begin{align}\label{eq:Ham_mol}
H=\sum_{ij,\sigma} t_{ij} \Big(c_{i\sigma}^\dagger c_{j\sigma}+c_{j\sigma}^\dagger c_{i\sigma} \Big),
\end{align}
where $c_{i\sigma}^\dagger (c_{i\sigma})$ creates (destroys) an electron on site $i$.

When the dimerization of the ET molecule pairs is strong, the
separation between bonding and antibonding bands is large and the
analysis of the low-energy tight-binding Hamiltonian can be reduced to
the two partially occupied antibonding bands. This case corresponds to
a half-filled anisotropic triangular lattice where the sites denote
the centers of the (ET)$_2$ dimers (centers of the two inner C-C bonds
on neighboring ET molecules):
\begin{align}\label{eq:Ham}
H=&\sum_{<ij>,\sigma} t\left(c_{i\sigma}^\dagger  c_{j\sigma} + c_{j\sigma}^\dagger c_{i \sigma}\right) + \sum_{[ij],\sigma} t' \left(c_{i\sigma}^\dagger c_{j\sigma} + c_{j\sigma}^\dagger c_{i \sigma}\right)
\end{align}
$t$ and $t'$ correspond to nearest- and next-nearest-neighbor hopping
contributions.  The hoppings between dimers (Eq.~\ref{eq:Ham}) can be
connected to the hoppings between molecules (Eq.~\ref{eq:Ham_mol})
using geometrical as well as analytical
considerations~\cite{Komatsu1996,Mori1998b,Tamura1991,Kandpal2009},
\begin{align}\label{dimer_t}
t\approx \frac{\vert t_2 \vert +\vert t_4 \vert}{2} \qquad \textnormal{and}\qquad  t' \approx \frac{\vert t_3 \vert }{2}.
\end{align}
Here $t_2$ and $t_4$ are the hoppings between one molecule and the
two closest molecules on the orthogonal oriented dimer. $t_3$ belongs
to the hopping between the closest ET molecules on neighboring equally
oriented dimers.

\begin{figure}[t]
\centering
\includegraphics[width=1\columnwidth]{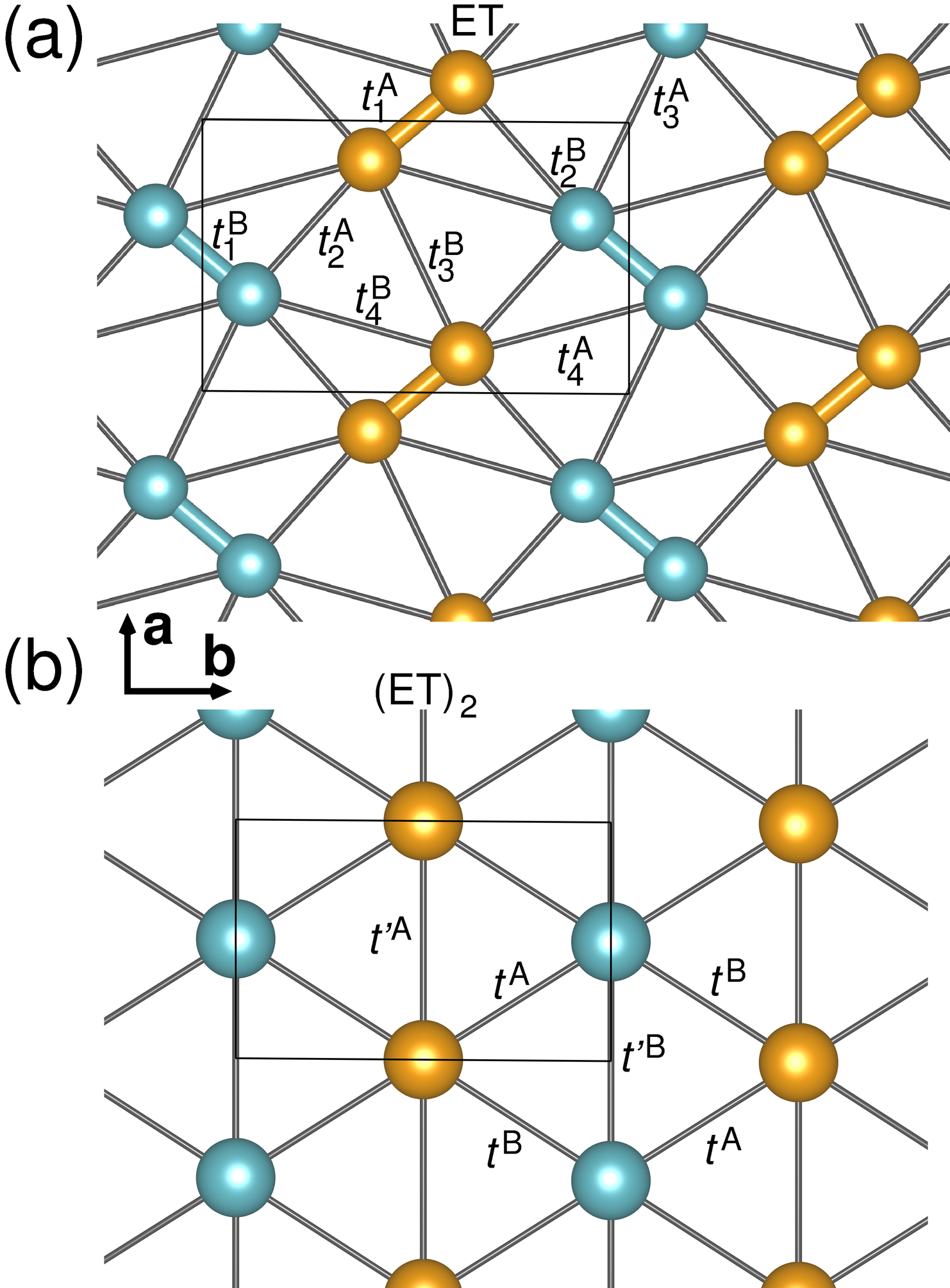}
\caption{(a) Illustration of the eight nearest neighbor hopping paths
  in the molecule model for the $\kappa$ layer of the dual-layered
  phase. (b) 
  Illustration of the tight-binding parameters of the dimer model.}
\label{fig:TBdual}
\end{figure}

In Table~\ref{tab:allti} we list the hopping parameters between ET
molecules (Eq.~\ref{eq:Ham_mol} and Fig.~\ref{fig:TBdual} (a)) in the
$\kappa$ layer of the three polymorphs {\comp}. Due to the presence of
two inequivalent (ET)$_2$ dimers in {\one} and {\two}, these phases
have twice as many TB parameters compared to the $\kappa$ phase.  For
the $\kappa$-only system, the approximate symmetry of $P\,2_1/c$ we
employ means that $\kappa$ molecules are all equivalent within the
layer but inequivalent between neighboring layers; however, as
differences between TB parameter sets are below $0.3$~meV, we report
only one of them in Table~\ref{tab:allti}.  Note, that the absolute
value of the TB parameters depends on the strength of the overlap of
the Wannier orbitals and is very sensitive to variations in the
distance and the orientation of the ET molecules, while the sign
originates from the phase factors of the TB Hamiltonian.  For
completion, we also show in Table~\ref{tab:alphadual} and
Table~\ref{tab:alphafour} the hopping parameters between ET molecules
in the $\alpha '$ layer for {\one} and {\two}, respectively. We observe
that the different charge distribution among the ET molecules in the
$\alpha'$ layers is a manifestation of the different crystal field
environment of the molecules as quantified by the onsite parameters
($t_0$) in Table~\ref{tab:alphadual} and Table~\ref{tab:alphafour}.
In Table~\ref{tab:ttprime}, the tight-binding parameters for the dimer
model (Eq.~\ref{eq:Ham} and Fig.~\ref{fig:TBdual} (b)) are listed.

\begin{table*}[htb]
\begin{ruledtabular}
\begin{tabular}{cc|cccccccccc}
system& &  $t_0^A$&  $t_0^B$&  $t_1^A$ & $t_1^B$ & $t_2^A$ & $t_2^B$ &
$t_3^A$ & $t_3^B$ & $t_4^A$ & $t_4^B$\\
\hline
\multirow{2}*{$\kappa$}&$t_i$\,(meV)&\multicolumn{2}{c}{-181.3}
&\multicolumn{2}{c}{168.0}
&\multicolumn{2}{c}{102.4}&\multicolumn{2}{c}{60.8}
&\multicolumn{2}{c}{33.4}  \\
&$d_i$\,(\AA) &\multicolumn{2}{c}{0} &\multicolumn{2}{c}{3.78}
&\multicolumn{2}{c}{5.55}&\multicolumn{2}{c}{6.79}
&\multicolumn{2}{c}{6.73}  \\
\hline
\multirow{2}*{\one}&$t_i$\,(meV) &
-166.7&-159.7&170.4&161.0&-98.8&-96.3&64.1&67.6&-38.1&-32.4 \\
&$d_i$\,(\AA) &0&0&3.74&3.77&5.60&5.65&6.64&6.66&6.84&6.85 \\
\hline
\multirow{2}*{\two}&$t_i$\,(meV)
&-164.4&-162.5&166.0&163.2&-98.7&-98.1&70.5&62.9&-35.0&-37.6\\
&$d_i$\,(\AA) &0&0&3.73&3.76&5.64&5.59&6.61&6.67&6.83&6.83\\
\end{tabular}
\end{ruledtabular}
\caption{Tight-binding parameters for the $\kappa$ layers of all three
phases. Distances between ET molecule centers are listed for
identification of hopping paths.}
\label{tab:allti}
\end{table*}

\begin{table*}[htb]
\begin{ruledtabular}
\begin{tabular}{cc|cccccccccccc}
system& &  $t_0^1$&  $t_0^2$&  $t_0^3$ & $t_0^4$ & $t_1^{13}$ & $t_1^{24}$ & $t_2^{12}$ & $t_2^{34}$ & $t_3^{14}$ & $t_3^{23}$ & $t_4^{14}$ & $t_4^{23}$\\
\hline 
\multirow{2}*{\one}&$t_i$\,(meV) & -55.9&-36.3&-189.1&-145.9&69.9&43.0&-9.5&-66.1&26.2&21.9&11.3 & 3.3 \\
&$d_i$\,(\AA) &0&0&0&0&4.21&4.21&6.61&6.61&7.81& 7.81 &7.86&7.86\\
\end{tabular}
\end{ruledtabular}
\caption{Tight-binding parameters for the $\alpha'$ layers of the dual-layered phase. Distances between ET molecule centers are listed for identification of hopping paths.}
\label{tab:alphadual}
\end{table*}
\begin{table*}[htb]
\begin{ruledtabular}
\begin{tabular}{cc|cccccccccccc}
system& &  $t_0^1$&  $t_0^2$&  $t_1$ & $t_2$ & $t_3$ & $t_4$ & $t_5$ & $t_6$ & $t_7$ & $t_8$ \\
\hline
\multirow{2}*{\two}&$t_i$\,(meV) &-90.2&-126.9&47.2&73.5&-35.0&-70.1&30.9&31.4&9.6&12.5\\
&$d_i$\,(\AA) &0&0&4.16&4.24&6.60&6.61&7.61&7.66&7.99&8.04\\
\end{tabular}
\end{ruledtabular}
\caption{Tight-binding parameters for the $\alpha'$ layers of the four-layered phase. Distances between ET molecule centers are listed for identification of hopping paths.}
\label{tab:alphafour}
\end{table*}

A detailed analysis on the hopping parameters for {\one} in
Table~\ref{tab:allti} shows that the tight-binding parameters
differentiate between the two inequivalent (ET)$_2$ dimers and reflect
the stripy charge transfer found for the $\alpha'$ layer and
illustrated in Fig.~\ref{fig:scheme}~(a). The more tightly bound dimers
(A) are above the stronger charged $\alpha'$ ET molecules with a $2/3$
hole; the less tightly bound dimers (B) are above the weakly charged
$\alpha'$ ET molecules with a $1/3$ hole.

In {\two} the strength of the dimerization as defined by the size of
the intradimer hoppings $t_1^A$ and $t_1^B$ is on average slightly smaller than in the
other two polymorphs.  Here, contrary to the {\one} case, the TB
parameters do not show the stripy pattern from {\one} since both the A
and B (ET)$_2$ dimers are aligned between the two distinctly charged ET molecules
in the $\alpha'$ layer (see Fig.~\ref{fig:scheme}~(b)).

\begin{figure}[t]
\centering
\includegraphics[width=1\columnwidth]{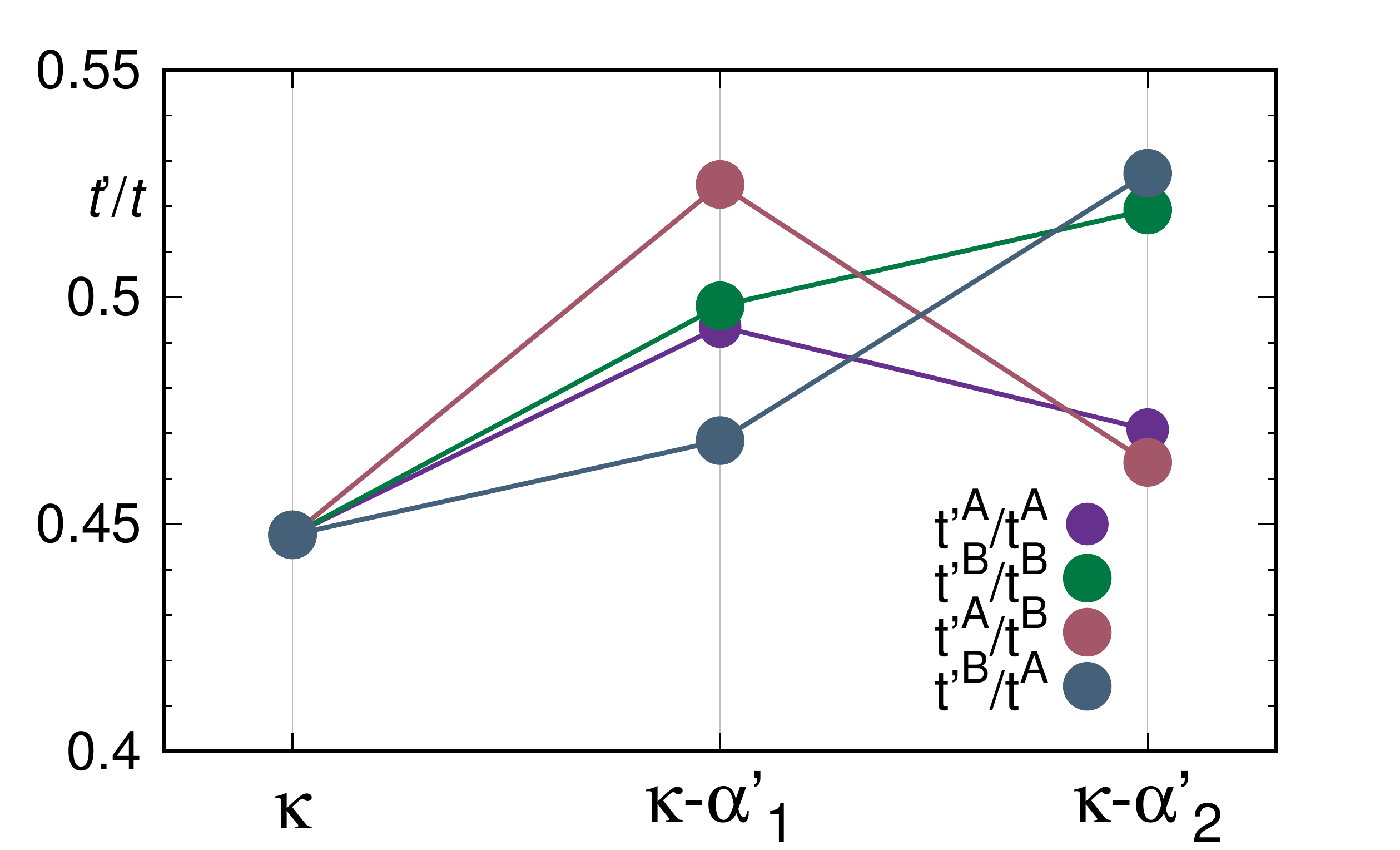}
\caption{Comparison of the $t'/t$ ratios of the three polymorphs which
  indicate the degree of frustration.}
\label{fig:frustration}
\end{figure}

In all three polymorphs, the dimer model estimate (Eq.~\ref{eq:Ham})
does not give such a good representation of the band structure (see
Appendix A) as, for
instance, in $\kappa$-(ET)$_2$Cu$_2$(CN)$_3$~\cite{Kandpal2009,Nakamura2009}. 
This may partly be due to the weak degree of dimerization in {\comp}, but
more importantly it reflects the fact that the basis defined by the
hoppings between dimer sites is not rich enough to correctly describe
details of the band structure and Fermi surface. Only by allowing the
larger basis defined by the hopping amplitudes between molecules can
all relevant details be captured. Keeping in mind that the dimer model 
only provides an approximate representation of the band structures,
we use it to  roughly compare the obtained trends to other previously examined 
$\kappa$ structures. The frustration is given by the ratio of the next-nearest 
hopping to the nearest hopping $\vert t'/t \vert $. 
Fig.~\ref{fig:frustration} shows that for the multilayered phases $|t'/t|\approx 0.5$, which is close to the value $0.58$ obtained
for $\kappa$-(ET)$_2$Cu(SCN)$_2$~\cite{Kandpal2009}, which is
superconducting at a temperature of 10.4 K. The low $T_c$ $\kappa$
phase is less frustrated and the frustration parameter of $0.45$ is
just slightly higher than the one calculated for the antiferromagnetic
insulator $\kappa$-(BEDT-TTF)$_2$Cu[N(CN)$_2$]Cl.

\begin{table}[htb]
\begin{ruledtabular}
\begin{tabular}{c|cccc}
system & $t^A$ & $t^B$ & $t'^A$ & $t'^B$\\
\hline 
{$\kappa$} &\multicolumn{2}{c}{67.9} &\multicolumn{2}{c}{30.4}  \\
\hline 
{\one} &-68.4 & -64.4 & 33.8 & 32.1 \\
\hline
 {\two}& -66.8 & -67.9 & 31.5 & 35.2  \\
\end{tabular}
\end{ruledtabular}
\caption{Table of triangular lattice Hamiltonian parameters $t$ and $t'$ for the three polymorphs, given in meV. }
\label{tab:ttprime}
\end{table}

\section{Discussion and Conclusions}

 Our microscopic analysis of the polymorph family {\comp} with
density functional theory and projective Wannier function
derived tight-binding Hamiltonians showed that, even though the
$\alpha'$ layer  in these materials is most probably insulating,
its electronic structure decisively influences
the behavior of the  conducting  $\kappa$ layer in {\one} and {\two}.
In particular, the TB hopping parameters in {\one} reflect the 
stripe-like charge ordered pattern shown by the ET molecules in the
$\alpha'$ layer. In {\two} there is no such pattern since the (ET)$_2$
dimers in {\two} are  always aligned between the two distinctly charged ET
molecules in the $\alpha'$ layer. This different charge arrangement
has its origin on the symmetry constraint imposed by $P\,2_1/n$ 
in {\two}, which leads to a different crystal field acting on the
ET molecules in the $\kappa$ layer.
Our tight-binding model parameters for the molecule-based model
(Eq.~\ref{eq:Ham_mol} and Table~\ref{tab:allti})
provide an adequate and reliable starting point for a many-body 
description of these systems in terms of a Hubbard-like Hamiltonian
where intramolecular and intermolecular Coulomb interaction terms
should be added to the tight-binding Hamiltonian.

In contrast, an analysis in terms of the simplified dimer model
(Eq.~\ref{eq:Ham} and Table~\ref{tab:ttprime}) shows that the
half-filled triangular lattice is not a very good starting point for
describing these materials. This is due to the relatively weak
dimerization of the (ET)$_2$ dimers and therefore the contribution of
the $\kappa$ bonding bands should not be neglected when describing the
electronic properties of these systems.  However, the information
obtained from the tight-binding hopping parameters in the dimer model
is still useful to roughly classify the degree of frustration in
{\comp} and have first hints for understanding the different
superconducting critical temperatures in these systems.  Our
comparison of the trends of the frustration parameters with earlier
studies shows that the $\kappa$-phase system lies in the range of
frustration where other $\kappa$ systems are antiferromagnetic
insulators, while the multi-layered {\one} and {\two} show a slightly
higher frustration degree as also observed in the superconductor
$\kappa$-(ET)$_2$Cu(SCN)$_2$.  Nevertheless, a detailed understanding
of the different critical temperatures requires a many-body analysis
of the molecular Hubbard-like model proposed here which is beyond the
scope of the present work.

Summarizing, in this work we investigated the electronic
properties of the charge-transfer compound (BEDT-TTF)$_2$Ag(CF$_3$)$_4$(TCE).
 We demonstrated and quantified the
importance of the $\alpha'$ layers with respect to the conducting
$\kappa$ layers   
 and suggested a
molecule-based model Hamiltonian to describe these systems. We hope
that this work will motivate other groups to investigate these
multi-layered materials which hold promise of increasing the
superconducting critical temperatures in organic charge-transfer
superconductors.

{\it{Acknowledgements}}

We thank D. Guterding and T. Kawamoto for helpful discussions.
We acknowledge support by the Deutsche Forschungsgemeinschaft through
Grant SFB/TR 49 as well as the Center for Scientific Computing (CSC) in
Frankfurt, Germany. 

\section{APPENDIX}

\subsection{Molecule versus dimer description}

\begin{figure}[H]
\centering
\includegraphics[width=\columnwidth]{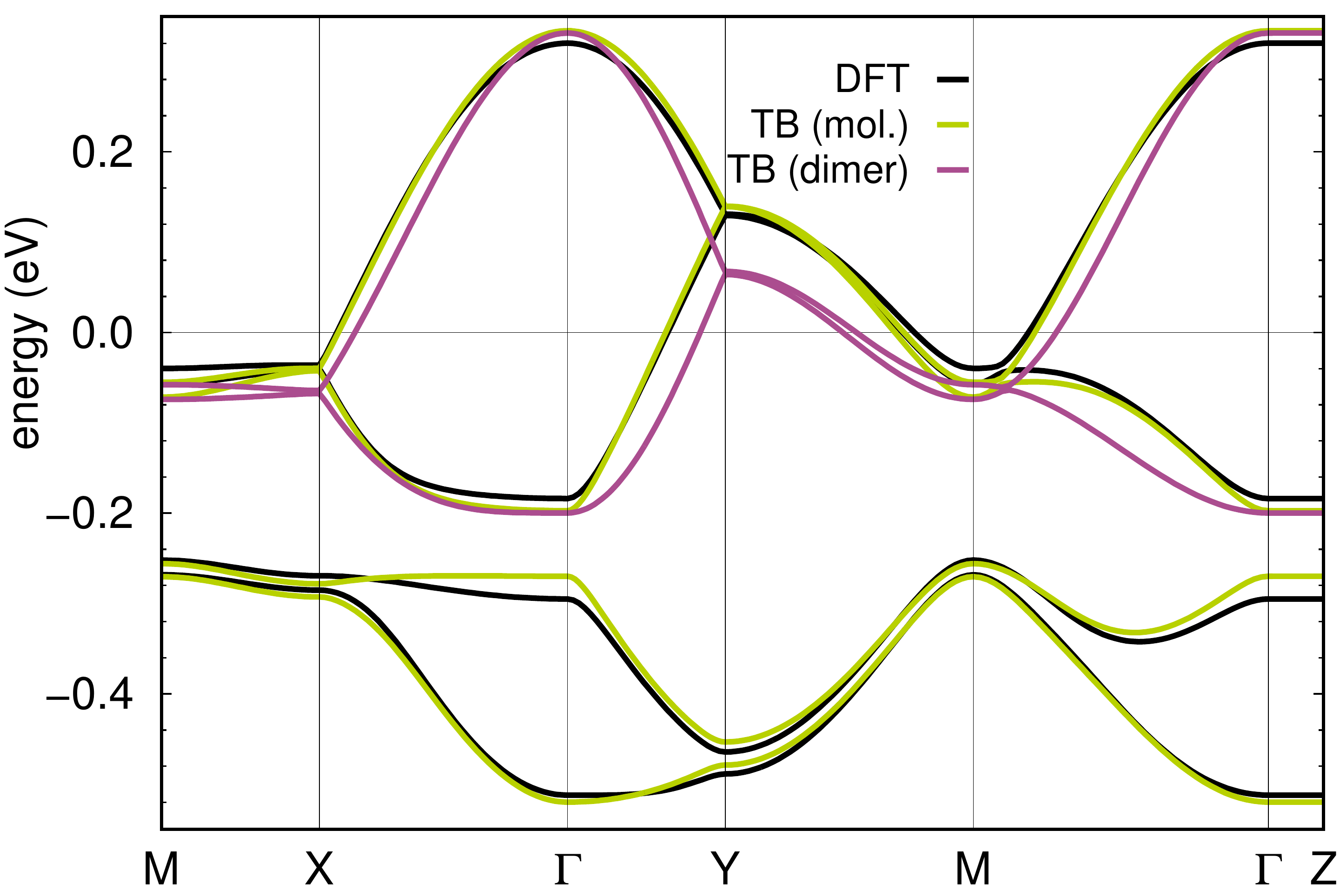}
\caption{Comparison of an eight parameter molecular TB model to the
  four parameter dimer TB model for the $\kappa$ layers of {\dual}.}
\label{fig:mol_dim}
\end{figure}

Fig.~\ref{fig:mol_dim} shows a comparison of the DFT calculated
$\kappa$ band structure in the {\one} phase to the molecular TB model
based on the eight parameters listed in Table~\ref{tab:allti}, and to
the dimer TB model based on the four parameters of
Table~\ref{tab:ttprime}. The parameters of the molecule model are
calculated using projective Wannier functions, while the dimer model
parameters are derived from them via geometrical relations,
Eq.~\eqref{dimer_t}. It is clear that the dimer model provides only a
rough approximation to the two half-filled $\kappa$ bands at the Fermi
level.

\subsection{Comparison of the $\kappa$ bands of the three polymorphs}

\begin{figure}[H]
\centering
\includegraphics[width=0.9\columnwidth]{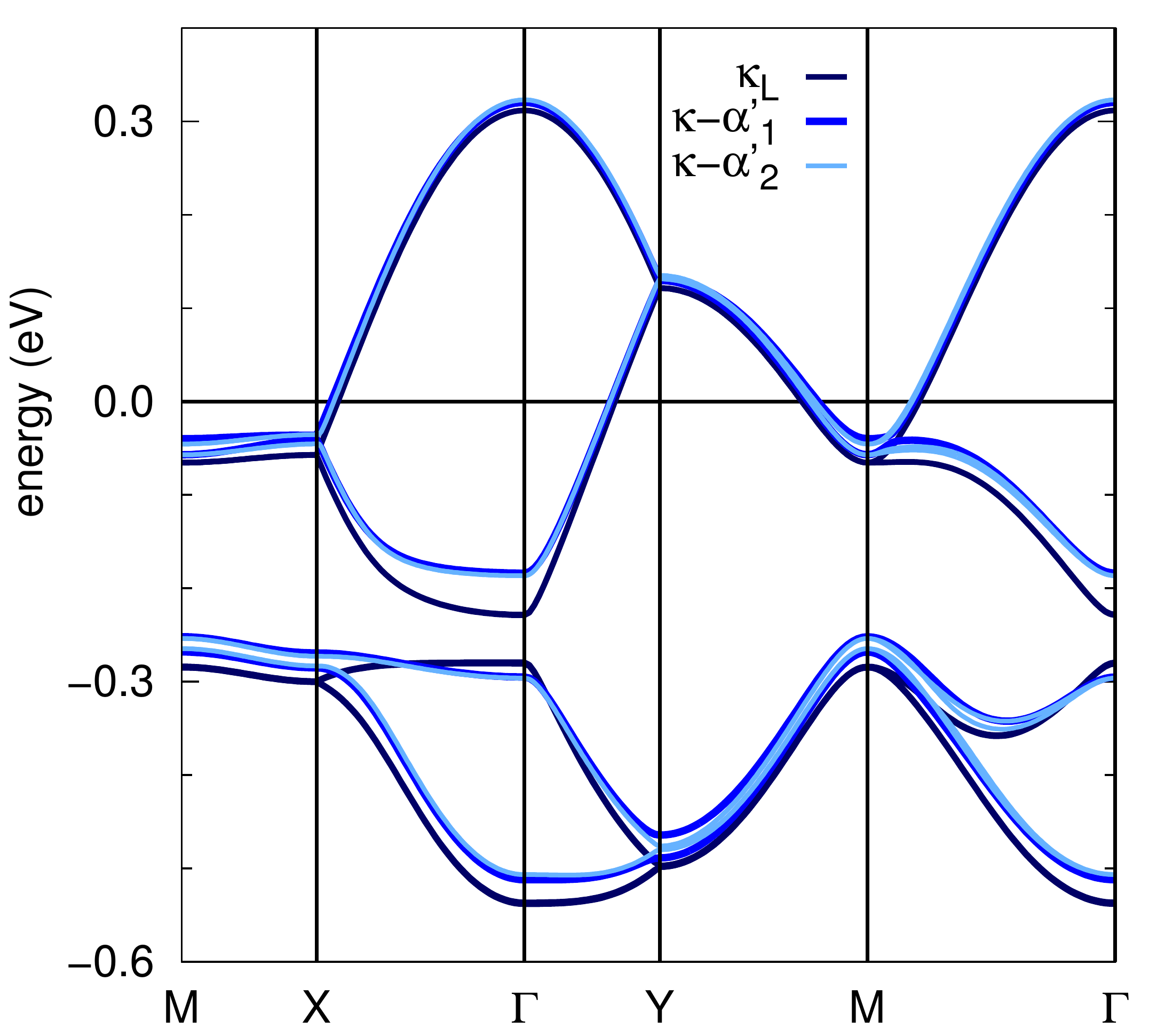}
\caption{Comparison of the $\kappa$ bands of the three polymorphs of
  {\comp} on an averaged path through the Brillouin zone (as the
  in-plane lattice parameters for the three compounds differ
  slightly).}
\label{fig:kappa_band_comp}
\end{figure}

In Fig.~\ref{fig:kappa_band_comp} we show  the $\kappa$ bands of the
three phases of (BEDT-TTF)$_2$Ag(CF$_3$)$_4$(TCE).

\end{document}